\documentclass[twocolumn,pre,superscriptaddress,showpacs,showkeys,nofootinbib,floatfix]{revtex4-1}

\usepackage[english]{babel}
\usepackage{amsmath,amssymb,graphicx}
\usepackage{algorithmic}
\usepackage{enumerate}
\usepackage{color}
\usepackage{soul}
\usepackage{enumitem}
\usepackage{tabto}
\usepackage{times}
\usepackage{textcomp}
\usepackage[pdftex]{hyperref}
\hypersetup{colorlinks=true,linkcolor=blue,citecolor=blue,urlcolor=blue}

\begin{document}

\title{Embedding Overhead Scaling of Optimization Problems in Quantum Annealing}

\author{Mario S. K\"onz}
\affiliation{Institute for Theoretical Physics, ETH Zurich, 8093 Zurich, Switzerland}

\author{Wolfgang Lechner}
\affiliation{Institute for Theoretical Physics,  University of Innsbruck, 6020 Innsbruck, Austria}

\author{Helmut G.~Katzgraber}
\thanks{The work of H.~G.~K.~was performed before joining Amazon Web Services}
\affiliation{Amazon Quantum Solutions Lab, Seattle, Washington 98170, USA}

\author{Matthias Troyer}
\affiliation{Microsoft Quantum, Microsoft, Redmond, Washington 98052, USA}
\affiliation{Institute for Theoretical Physics, ETH Zurich, 8093 Zurich, Switzerland}

\date{\today}
\begin{abstract}

In order to treat all-to-all connected quadratic binary optimization problems (QUBO) with hardware quantum annealers, an embedding of the original problem is required due to the sparsity of the hardware's topology. Embedding fully-connected graphs---typically found in industrial applications--- incurs a quadratic space overhead and thus  a significant overhead in the time to solution. Here we investigate this embedding penalty of established planar embedding schemes such as square lattice embedding, embedding on a Chimera lattice, and the Lechner-Hauke-Zoller scheme using simulated quantum annealing on classical hardware. Large-scale quantum Monte Carlo simulation suggest a polynomial time-to-solution overhead. Our results demonstrate that standard analog quantum annealing hardware is at a disadvantage in comparison to classical digital annealers, as well as gate-model quantum annealers and could also serve as benchmark for improvements of the standard quantum annealing protocol.

\end{abstract}


\maketitle

\section{INTRODUCTION}

The availability of commercial quantum annealing devices \cite{johnson:11,dickson:13,bunyk:14} has revolutionized (quantum) optimization across many disciplines. The fruitful race \cite{pudenz:13,smith:13,boixo:13a,albash:15a,ronnow:14a,katzgraber:14,lanting:14,santra:14,shin:14,vinci:14,boixo:14,albash:15,albash:15a,albash:18,katzgraber:15,martin-mayor:15a,pudenz:15,hen:15a,venturelli:15a,vinci:15,zhu:16,mandra:16b,mandra:17a,mandra:18} between quantum optimization using quantum annealing
\cite{finnila:94,kadowaki:98,brooke:99,farhi:01,santoro:02,das:05,santoro:06,das:08,morita:08,hauke:20} and classical algorithms on traditional CMOS hardware has resulted in multiple new ways to solve hard problems of practical importance previously thought to be intractable.

Despite overcoming formidable engineering challenges in building superconducting quantum optimization machines, some fundamental limitations will remain as seemingly insurmountable challenges for years to come \cite{katzgraber:18a}. Leaving analog noise aside, both {\em mapping} (binary) optimization problems onto $2$-local Hamiltonians, as well as the {\em embedding} of the resulting $2$-local Hamiltonian onto the hardwired topology of the quantum annealer, represent potentially large bottlenecks for solving application problems that typically do not perfectly fit the hardware layout. Already the mapping of general Hamiltonians onto a $2$-local model, such as a quadratic unconstrained binary optimization problem (QUBO), can result in a large  overhead in the number of variables. In addition, the lack of reliable arbitrary long-range couplers between qubits in superconducting systems results in a quasiplanar layout of quantum annealing devices. While newer generations of superconducting quantum annealing devices have increased the qubit connectivity, the underlying topology remains local and therefore unable to natively accommodate long-range couplers between variables, which are often present in applications.

Such intrinsic limitations of analog computing devices have played a major role in the shift to programmable digital computers in classical computing. Digital approaches, either on classical hardware or on future gate-model quantum computers, circumvent the embedding problem because the programmable logic of these devices allows for arbitrary Hamiltonians on arbitrary graphs. While scalable gate-model quantum computers are not yet available, quantum-inspired optimization methods \cite{mandra:16b} digitally implemented on classical hardware have gained traction to solve application problems \cite{perdomo:17y,tsukamoto:17,matsubara:17,aramon:18,hamerly:18x}. For current superconducting quantum annealing hardware, however, highly connected $2$-local {\em logical} problems with $N$ variables need to be embedded onto the sparse-connectivity hardware graph of ${\cal O}(N^2)$ {\em physical} qubits. The additional degrees of freedom of the embedded problem need to be restricted using additional constraints in form of additional coupling terms. While this space overhead (which, incidentally, can sometimes be used for error-correction schemes \cite{pudenz:13,pudenz:15}) has been well studied, here we demonstrate that there is an even more significant time overhead in time to solution \cite{ronnow:14a,mandra:16b}.

\begin{figure}[t]
\centering
\includegraphics[width=1\columnwidth]{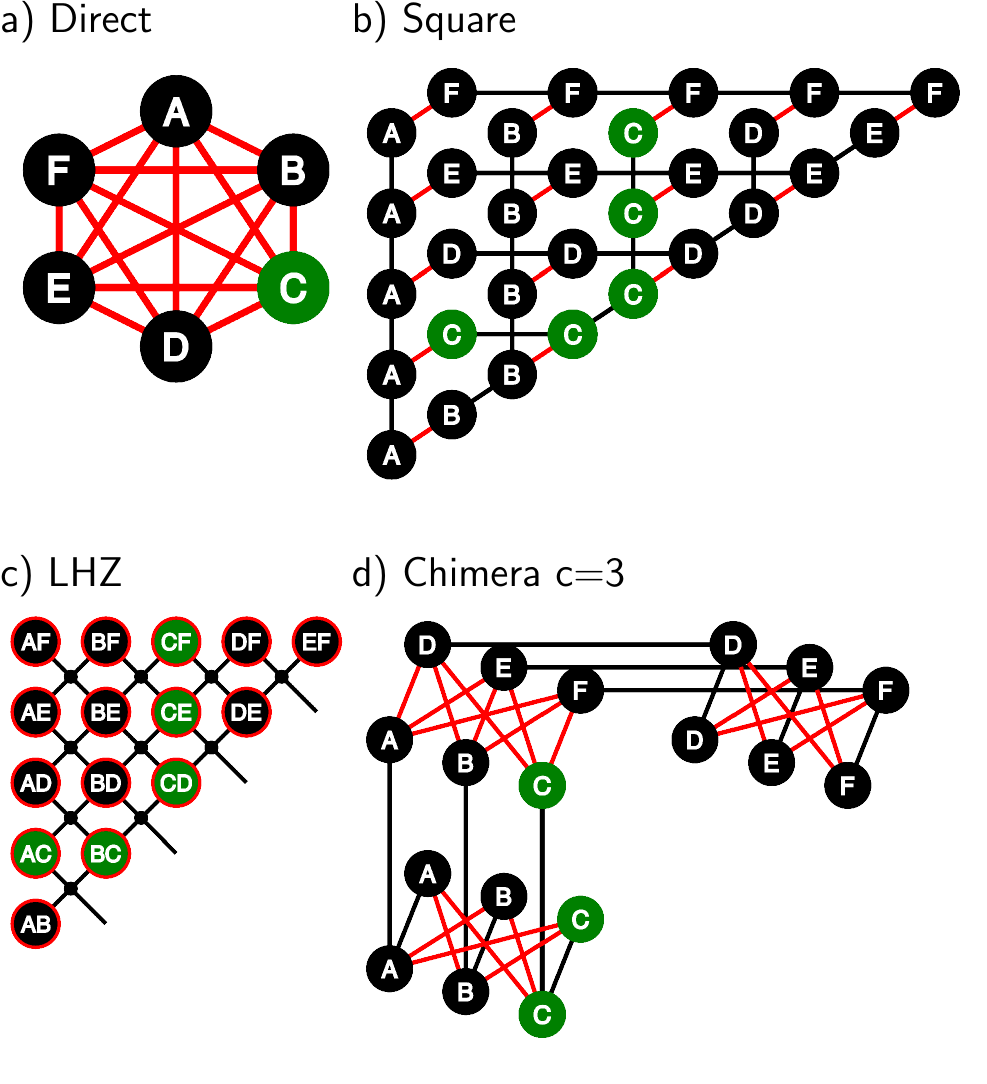}
\caption{\label{f_mappings}
(a) Direct optimization. The fully connected problem with six binary variables (circles) labeled ``A--F.'' Couplers between the variables are represented as solid (red) lines. (b) The square lattice. The logical variable C is represented by a chain of $N-1$ logical qubits. Constraints are represented by black (dark) lines. These chains are laid out such that every variable chain can connect to every other variable chain, shown by the red (light) lines. (c) The LHZ scheme. In the PAQO embedding, a physical variable $\alpha_{i,j}$ is the product of two logical binary variables $\sigma_i \sigma_j$. The logical couplers correspond to physical local biases (red). Constraints are enforced with $4$-local interactions (black crosses) between neighboring spins. (d) The chimera lattice, $c=3$. A $K_{3,3}$ Chimera lattice consisting of a square-embedding-like chain structure (black or dark lines) combined with bipartite graph cells. These cells contain the logical couplers (red or light lines).
}
\end{figure}

Specifically, in this work, we focus on three embedding schemes: embedding on a square lattice, embedding on a chimera lattice, and the parity adiabatic quantum optimization (PAQO) or Lechner-Hauke-Zoller (LHZ) scheme. As an example, we illustrate these embedding approaches using the fully connected six-variable binary problem depicted in Fig.~\ref{f_mappings}(a). Our focus in this work is on the {\em embedding overhead}.  As such, we do not discuss the {\em locality reduction overhead} introduced by converting a binary $k$-local problem to a $2$-local. Also, without loss of generality, we assume that the problem does not have local biases (i.e., a field or $1$-local term), because these can always be implemented as $2$-local terms with a fixed auxiliary variable. For square and chimera embedding we use majority decoding, while belief propagation \cite{pastawski:16} is used for the PAQO scheme.

The paper is structured as follows. In Sec.~\ref{sec:emb_scheme}, we discuss the different embedding strategies analyzed in this work, and then, in Sec.~\ref{sec:benchmarking}, we outline the technical details of the benchmarking. Section \ref{sec:qmc} shows our main results, followed by Sec.~\ref{sec:emb} on embedding penalties. We summarize our findings in Sec.~\ref{sec:conclusion}. Appendices \ref{apx:paqo_constr} and \ref{apx:avoid} contain two technical notes about PAQO constraints and avoided level crossings.

\section{DETAILS ON EMBEDDING SCHEMES}
\label{sec:emb_scheme}

In this section we describe in detail the different embedding schemes, as well as the corresponding decoding strategies. In general, we aim to optimize a binary quadratic optimization problem on a complete graph, i.e.,
\begin{equation}
    {\mathcal H} = \sum_{i < j} J_{ij} \sigma_i \sigma_j ,
    \label{eq:ham}
\end{equation}
where the variables $\sigma_{i} \in \{\pm 1\}$ and the problem is specified by the values of the couplers $J_{ij}$. Note that in this work we do not study linear terms in Eq.~\eqref{eq:ham} without loss of generality.

\subsection{Embedding on a square lattice}

Square-lattice embedding reduces an all-to-all logical system to a physical system connected only locally. This is achieved by representing each logical variable as a chain of strongly coupled physical qubits. Figure \ref{f_mappings}(b) shows the square embedding for the example presented in Fig.~\ref{f_mappings}(a).  Each logical variable is represented by a chain of qubits of length $N-1$.  With a careful chain layout, each chain of physical qubits has a coupler to every other chain, where the logical couplings (red, light color) are located. The black (dark) couplers along the chains are the constraints that ensure that all variables on the chain have the same value, i.e., either all up or all down. If the logical problem has a special structure or limited connectivity, the embedding of the logical problem can be optimized to use as few physical qubits as possible with the minor embedding \cite{choi:08}. An optimal optimization, however, poses the risk of being computationally more expensive to solve than the actual optimization problem itself. Since we use heavily connected as well as all-to-all-connected problems, use of the minor embedding yields no benefits. We choose the constraints for each logical variable as
\begin{equation}
C_i(\omega, \gamma) =  \omega + \gamma \sum_{j\neq i} |J_{ij}|,
\label{eq:gamma}
\end{equation}
where $\omega$ is a constant determining a fixed base constraint, while $\gamma$ is the so-called ``sum constraint'' that is multiplied by the sum of the absolute value of the couplers $J_{ij}$ involving variable $i$. This allows us to constrain variables with more or larger couplings more strongly than variables with fewer or smaller couplings, therefore ensuring that the system is not overconstrained. In general, one wants to use constraints that are as small as possible because large penalty terms tend to affect the performance of quantum annealing hardware \cite{comment:fn1}. However, if the constraints (penalties) are not strong enough, the chains that link physical variables to create logical counterparts have the potential to ``break'', thus losing the logical information about the problem. If the decoding strategy can decode errors (broken chains), we might even deliberately choose weak constraints and fix the errors later, as this is beneficial for solving the problems faster. The chain length scales linearly with the number of variables. Similarly, if $\gamma\neq0$, the constraints also scale linearly with the number of variables, as an increased number of variables typically increases the value of the sum in Eq.~\eqref{eq:gamma}.

Once the logical problem is embedded in the physical system, the optimization procedure is executed. The physical solution then needs to be decoded to obtain the solution for the underlying logical system. If no constraints are broken, there is a trivial mapping from the chains to the logical variables. If constraints are broken, there are many decoding strategies for square-lattice embedding. We choose a straightforward and computationally inexpensive strategy here. For every chain, \emph{majority voting} determines the value of the logical variable, depending on the value of the majority of physical variables in a chain.

\subsection{Embedding on a chimera lattice}

Chimera-lattice embedding \cite{bunyk:14}, which is used in the D-Wave 2000Q device, uses a variant of the aforementioned square-lattice embedding. The chimera lattice consists of bipartite unit cells of size $2c$ for some fixed integer $c$. For the D-Wave 2000Q, $c = 4$. A logical variable is represented by a chain of physical qubits but it takes fewer qubits to represent the variable due to the higher connectivity in a single $K_{4,4}$ cell.  Figure \ref{f_mappings}(d) shows the embedding on a $c=3$ chimera lattice for the example shown in Fig.~\ref{f_mappings}(a). The cells at the bottom and on the right contain the same $c$ qubits, while all the other cells (in this example, only one) contain $2c$ unique qubits. As for embedding on the square lattice, we use majority voting for decoding. Note that the embedding strategy outlined here can also be used for the new generation of D-Wave devices, i.e., the D-Wave Advantage with the new Pegasus topology. Because the connectivity is higher, shorter chain lengths are needed. However, the lattice remains quasiplanar.

\subsection{Parity adiabatic quantum optimization}

An alternative way to map the logical problem has recently been proposed in Ref.~\cite{lechner:15}. The PAQO or LHZ scheme encodes the logical problem differently to square- and chimera-lattice embedding. Instead of having the notion of logical variables in the physical system, each physical variable encodes the product between two logical variables $\alpha_{i,j}=\sigma_i\sigma_j$, i.e., stores if they are equal or opposite. The main advantage of this approach is the mapping from logical couplings to physical local biases, which can be controlled better than two-qubit physical couplers. Figure \ref{f_mappings}(c) shows the PAQO embedding for the example in Fig.~\ref{f_mappings}(a). While the locality of the Hamiltonian goes from quadratic to quartic, a fully connected nonplanar graph turns into a square lattice. This embedding will be of considerable relevance once four-way couplers become available on quantum annealers.

The constraints are modeled as $4$-local terms \cite{comment:fn3} between four variables that form a tile. Fixed variables that are positive can be introduced at the lower-right edge to complete tiles [not shown in Fig.~\ref{f_mappings}(c)]. Because every logical variable appears twice in a constraint term, it must be $1$, i.e., for the example in Fig.~\ref{f_mappings}(a),
\begin{equation}
\alpha_{\rm AH} \alpha_{\rm AG} \alpha_{\rm BH} \alpha_{\rm BG} = \sigma_{\rm A} \sigma_{\rm H} \sigma_{\rm A} \sigma_{\rm G}  \sigma_{\rm  B} \sigma_{\rm H} \sigma_{\rm B} \sigma_{\rm G} = 1,
\end{equation}
where $\alpha$ denotes physical variables (qubits) and $\sigma$ denotes logical variables.  This that means every tile needs an even number of negative physical variables. The Hamiltonian is therefore defined as

\begin{equation}
    {\mathcal H_{\text{PAQO}}} = \sum_{i < j} J_{ij} \alpha_{i,j} + \sum_{l} C_{l}~\alpha_{\nwarrow} \alpha_{\swarrow}\alpha_{\searrow}\alpha_{\nearrow}
\end{equation}
where $l$ iterates over all $4$-local constraints that have one physical spin in each corner and $C_l(\omega,\gamma)=\omega+\gamma~(J_{\nwarrow} + J_{\swarrow}+J_{\searrow}+J_{\nearrow})$, analogous to Eq.~\ref{eq:gamma}.

In general, the scaling of the constraints depends on the problem class and can be analytically understood for the ferromagnetic, antiferromagnetic and spin-glass cases with exponents between $0.5$-$2$ \cite{lanthaler2020minimal}. Here, we choose to scale the constraints according to the worst-case scenario, i.e., $\propto N^2$, which is motivated in Appendix~\ref{apx:paqo_constr}. An alternative is to set the constraints, perform the optimization, and then verify if any of the constraints are broken. If so, increase the value and iterate. This, however, can be computationally costly.

The PAQO embedding shares similarities with a low-density parity code, which can be decoded using \emph{belief propagation}, as also proposed in Ref.~\cite{pastawski:16}.  The original choice of constraints is not unique in any way; there are many combinations of physical variables that form loops, meaning that each logical variable occurs an even number of times leading the expression to be $1$. Belief propagation uses this fact and creates, for example, all possible three-variable loops, e.g., $\alpha_{AB}\alpha_{AC}\alpha_{BC}=1$ and determines which of the physical variables are most likely to be wrong in a broken constraint.  While the random bit-flip error resilience of the PAQO scheme might be very useful in an implementation, it cannot correct for single broken constraints as shown in Fig.~\ref{f_paqo_theo}.  Belief propagation then decodes the wrong logical state if started from Fig.~\ref{f_paqo_theo}(e) compared to Fig.~\ref{f_paqo_theo}(b). Hence, reducing the constraints is undesirable.

\section{METHODS}
\label{sec:benchmarking}

\begin{figure}[h!]
\centering
\includegraphics[width=1\columnwidth]{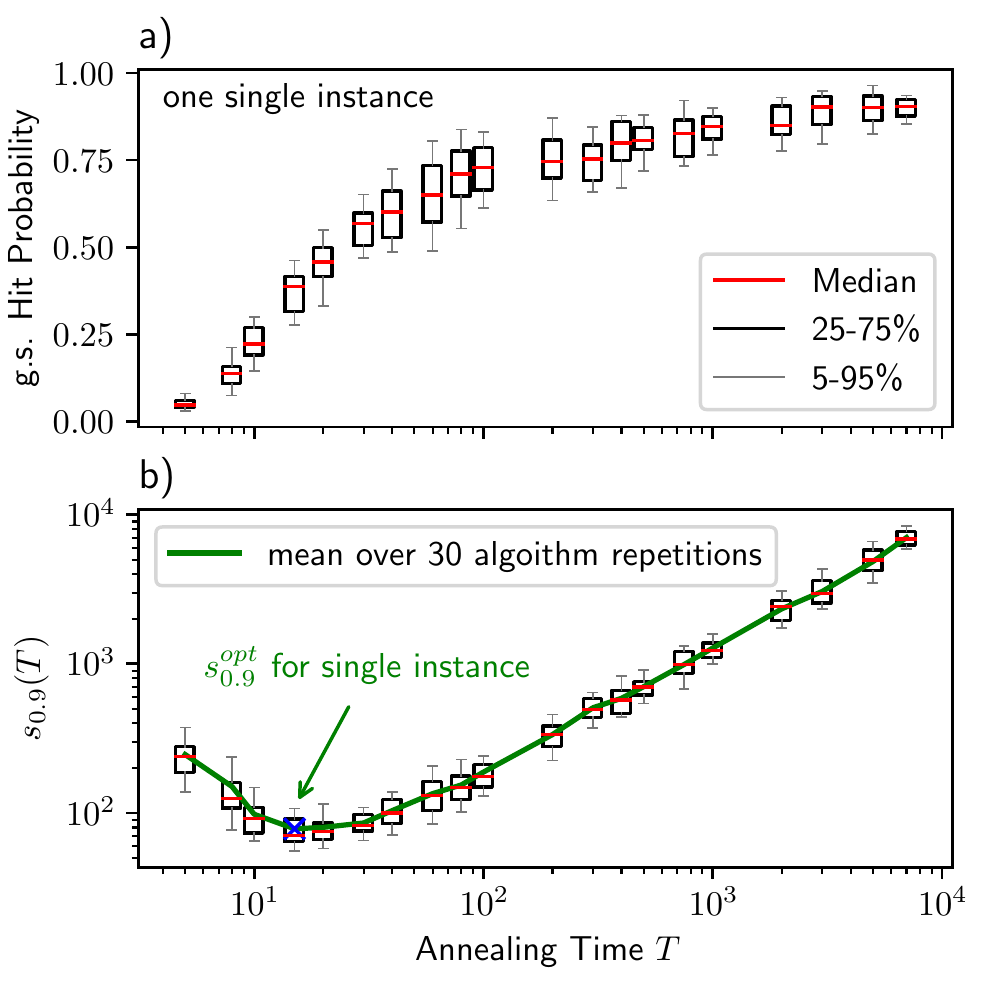}
\caption{\label{f_tts_vs_T_single}
(a) The ground-state hit probability (g.s.-Hit) for various annealing times $T$ using simulated quantum annealing. The slower we anneal, the more likely we are to find the ground state. The specific instance used in this example is a MaxCut instance of size $N = 25$ and density $p=0.3$. For each annealing time, we repeat the algorithm $30$ times. The box plot shows the first to third quartile, while the whiskers represent 5\% and 95\% of the measured data. (b) The time to solution (with a probability of 90\%)  versus the annealing time $T$. Because the ground-state hit probability is not a linear function, it is faster to repeat short-annealing-time runs than to invest in one long run. Therefore, we measure the mean and not the median. Both panels have the same horizontal axis.
}
\end{figure}

We compare the different embedding schemes using the time to solution (TTS) \cite{ronnow:14a} for quantum annealing with the goal of determining the cost overhead compared to a direct optimization of the logical problem. The TTS $s$ is defined via
\begin{equation}
    s_{p_\text{tar}}(T)= T  r_{p_\text{tar}}(T),
\end{equation}
where
\begin{equation}
r_{p_\text{tar}}(T)=\frac{\log[1-p_\text{tar}]}{\log[1-p_{\text{g.s.}}(T)]}
\end{equation}
is the number of repetitions needed to obtain a given target success probability $p_\text{tar}$, $T$ is the anneal time, and $p_{\text{g.s.}}(T)$ is the probability that the ground state is found. There are two possible types of ground states that interest us, the physical one and, more importantly, the logical one. It is valuable to investigate the physical ground state, because it can be decoded trivially. The result can then be compared to that obtained by the decoder to check its performance. We use a simple grid search to determine the optimal parameters and thus extract the true scaling. This works well because $s_{p_\text{tar}}(T)$ is typically a convex function, as shown in Fig. \ref{f_tts_vs_T_single}(b). To obtain the optimal $s^{\rm opt}$ that describes the actual scaling, the TTS is minimized with respect to the anneal time $T$, i.e.,
\begin{equation}
s_{p_\text{tar}}^{\rm opt}=\min_{T}[s_{p_\text{tar}}(T)].
\end{equation}
To keep the notation concise, we use $s_P=s_{90\%}^{\rm opt,physical}$ for the optimal time to solution of the physical ground state and $s_L=s_{90\%}^{\rm opt,logical}$ for the logical ground state after the decoder is run.

As a benchmark problem and a proxy for real-world applications, we use two instance types. First, we use unweighted MaxCut instances characterized by their connectivity density $p$, where $p=1$ ($100\%$) corresponds to an all-to-all-connected graph. Given a graph of size $N$ and density $p$, we generate instances with $N$ vertices and $N_e=p [N (N-1)]/2$ edges.  This number is then rounded to the nearest integer. The seed $s$ initializes the random number generator that draws $N_e$ random edges from the set of all possible edges. All edges have an antiferromagnetic weight, i.e., $J=-1$. The second problem family class consists of all to all connected ($p=1$) instances with weights drawn from a Gaussian distribution, such that there are ferromagnetic and antiferromagnetic weights, i.e., a Gaussian Sherrington-Kirkpatrick \cite{sherrington:75} spin glass.

We simulate $31$ different logical problems with size $N = 5$ -- $35$ variables. For each system size $N$, we generate $30$ random instances of three types: $p = 0.3$ MaxCut, $p = 0.5$ MaxCut, and fully connected Gaussian spin glasses with mean $\mu=0$ and variance $\sigma=1$. Using simulated quantum annealing, we study the logical problem directly, the problem embedded in a square lattice, the problem embedded on a chimera lattice and the problem encoded in the PAQO scheme. For each instance, we use $30$ repetitions of the transverse-field-simulated quantum annealing \cite{santoro:02,heim:15,isakov:16,mazzola:17} algorithm with $29$ different annealing times between $5$ and $28 000$ Monte Carlo sweeps. Furthermore, we use $1024$ Trotter slices and an inverse temperature of $\beta=1024$. The transverse field (annealing schedule) is varied from $0.5$ to $0.001$ in a square manner, with the steep edge at the beginning. Multiplying all dimensions yields approximately $1.1 \times 10^7$ simulations to be performed. To reduce the search space, we drop selected parameter combinations, e.g., a small logical problem size with a long annealing time and vice versa. This, in turn, reduces the search space to approximately $9$ million simulations. Finally, for square- and cimera-lattice embedding we use $\omega=0$ and $\gamma=1.1$ in Eq.~\eqref{eq:gamma}, while for the PAQO scheme we use $\omega=N^2/50$ and $\gamma=1.1$ for the constraints.

As an example, Fig.~\ref{f_tts_vs_T_single}(a) shows the ground-state (g.s.) hit probability for a single specific instance and different annealing times. By counting the Trotter slices that have the ground-state energy, we can compute the ground-state hit probability, because a measurement corresponds to choosing one random Trotter slice. We then convert this quantity to $s_{0.9}$ and display the result in Fig.~\ref{f_tts_vs_T_single}(b) where $s_{0.9}^{opt}$ is marked by an arrow.

\begin{figure}[h!]
\centering
\includegraphics[width=1\columnwidth]{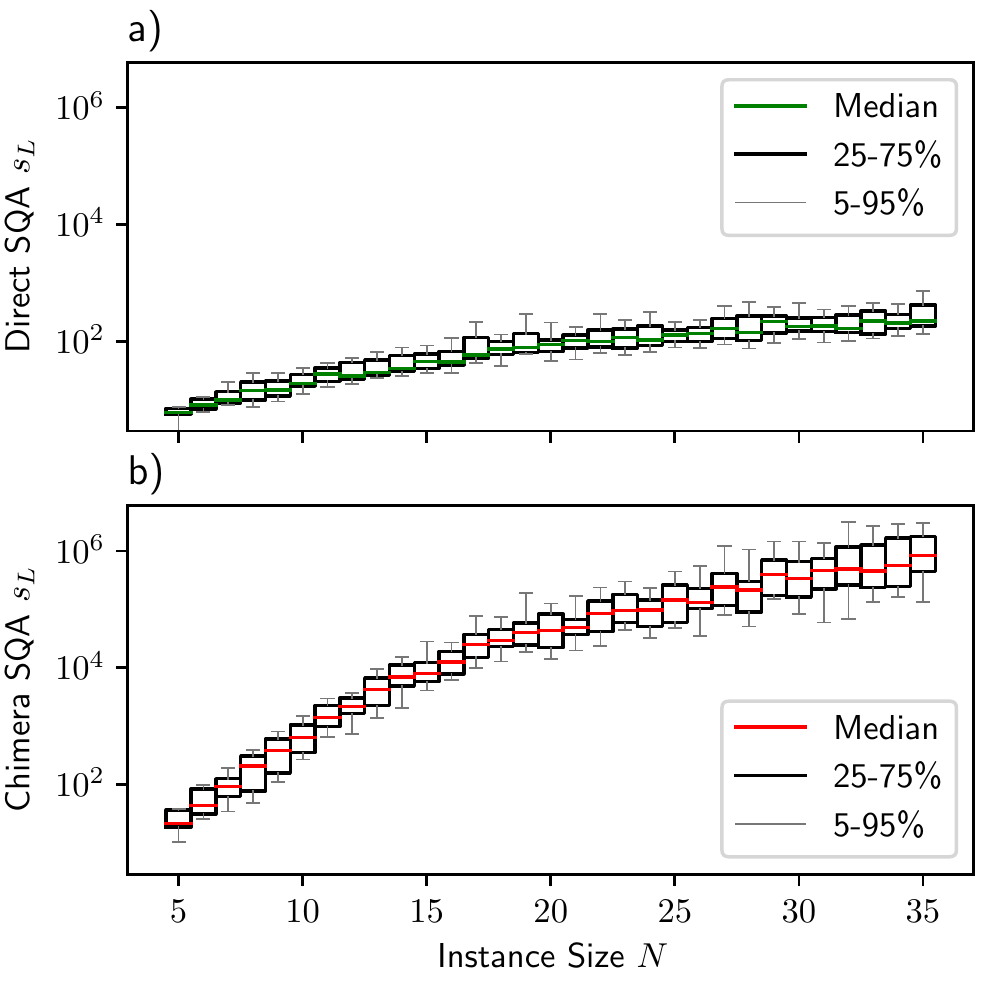}
\caption{\label{f_vert_vs_tts}
(a) All $s_{0.9}^{opt}$ for a physical problem using simulated quantum annealing (SQA) [one highlighted in Fig.~\ref{f_tts_vs_T_single}(b) with a green arrow] of all instances, with density $p=0.5$ sorted by size in a linear-log box plot. (b) The same as (a), but for the problems embedded in the chimera lattice. Both panels have the same horizontal axis.
}
\end{figure}

Following the same procedure, we calculate $s_{0.9}^{opt}$ for {\em every} individual instance. It is quite possible that different MaxCut instances of the same size and density display their $s_{0.9}^{opt}$ at different $T$. Figure \ref{f_vert_vs_tts}(a) shows these $s_{0.9}^{opt}$ versus different instance sizes. As an example, we show data for the logical problems embedded onto the chimera lattice.

To estimate the embedding overhead, we compare the data in Fig.~\ref{f_vert_vs_tts}(b) (post chimera embedding) with Fig.~\ref{f_vert_vs_tts}(a) (direct optimization of the logical problem) by plotting $s_{0.9}^{opt}$ in a scatter plot for the different embedding schemes in Sec.~\ref{sec:qmc}.

\section{QUANTUM MONTE CARLO RESULTS}
\label{sec:qmc}

\begin{figure}[h!]
\centering
\includegraphics[width=1\columnwidth]{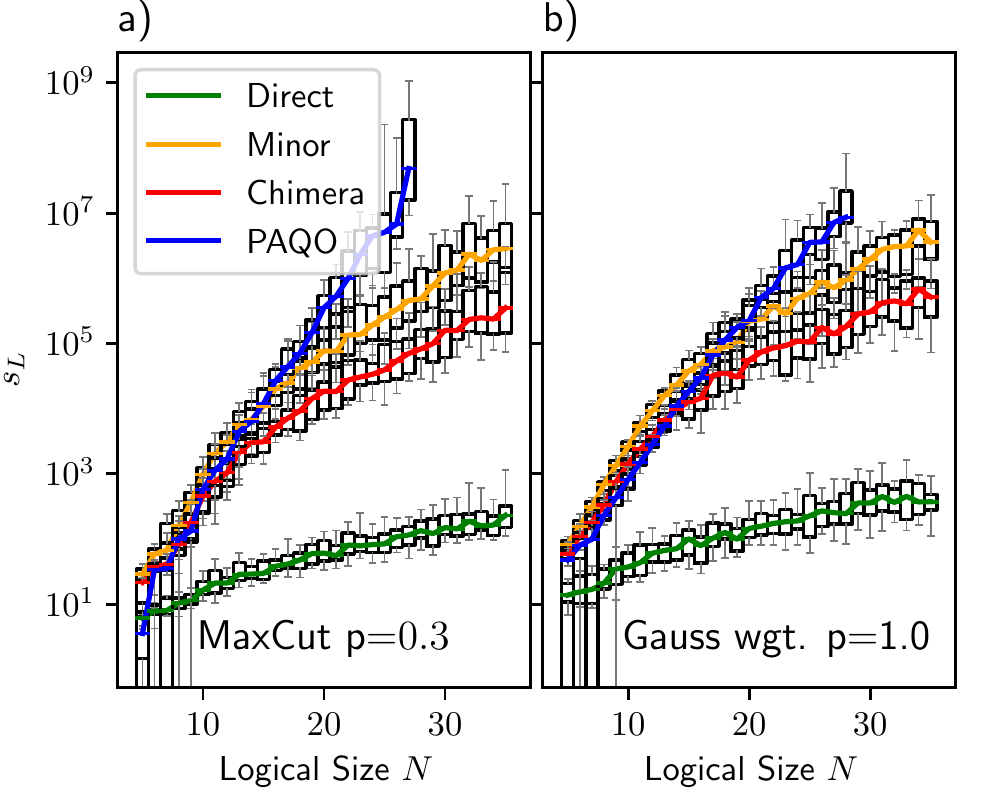}
\caption{\label{f_vert_vs_tts_box}
$s_L$ vs the logical system size $N$ for (a) MaxCut instances with $p = 0.3$ and (b) Gaussian spin glasses. The data shown are for the three embedding schemes, as well as a direct simulation of the logical problem (bottom data set). The $s_L$ for each instance is determined by averaging over $30$ random starts of the algorithm. The reason for the absence of the largest system size for the PAQO scheme is the inability to find the ground state, likely due to insufficient constraint scaling. Both panels have the same vertical axis.
}
\end{figure}

\begin{figure}[h!]
\centering
\includegraphics[width=1\columnwidth]{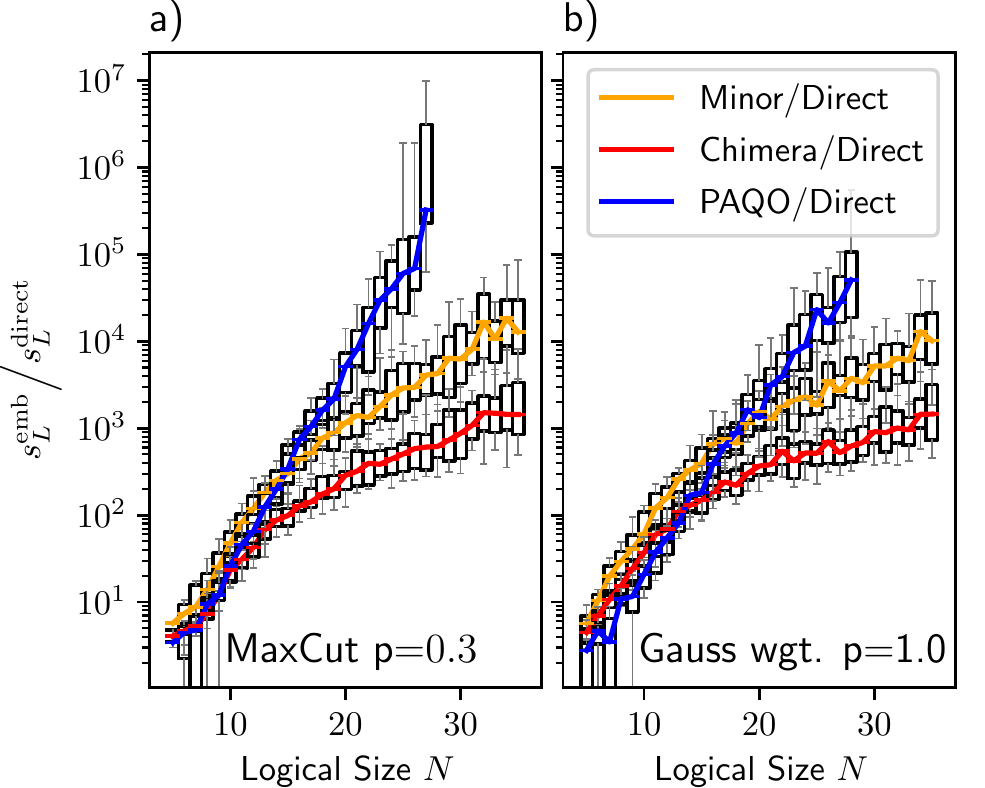}
\caption{\label{f_vert_vs_tts_div_box}
The same data as in Fig.~\ref{f_vert_vs_tts_box} but divided by $s_L^{\rm direct}$. Both panels have the same vertical axis.
}
\end{figure}

Figure \ref{f_vert_vs_tts_box} shows the $s_L$ of the direct simulation and the three embedding schemes versus the logical system size $N$. The data clearly show a time-based penalty due to the different embedding schemes over the direct simulation of the logical problem. To better quantify the embedding overhead, we study the ratio between the TTS of the logical and physical problems. The data are shown in Fig.~\ref{f_vert_vs_tts_div_box} as an instance-by-instance scatter plot. The data have an approximately linear trend (log-log plot), indicating a polynomial overhead in the time to solution.

\begin{figure}[h!]
\centering
\includegraphics[width=1\columnwidth]{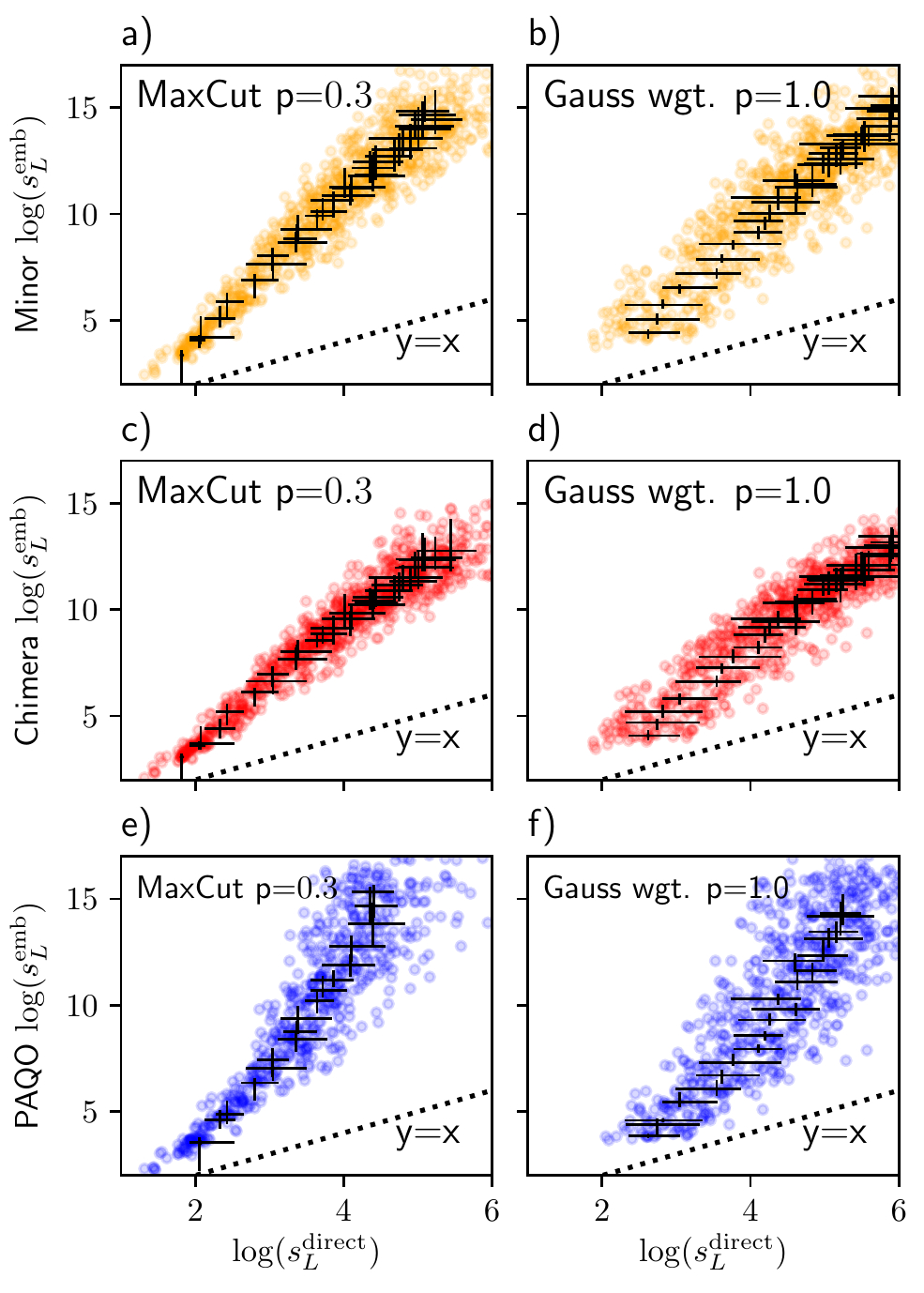}
\caption{\label{f_tts_vs_tts_separate_scatter}
The same data as in Fig.~\ref{f_vert_vs_tts_box} but displayed with $\log(s_L)$ embedded versus direct optimization of every independent instance as a scatter plot. The black crosses show the first to third quartiles for each logical size. The linear scaling in a log-log plot indicates a polynomial scaling. All panels share the same vertical and horizontal axes.
}
\end{figure}

\begin{figure}[h!]
\centering
\includegraphics[width=1\columnwidth]{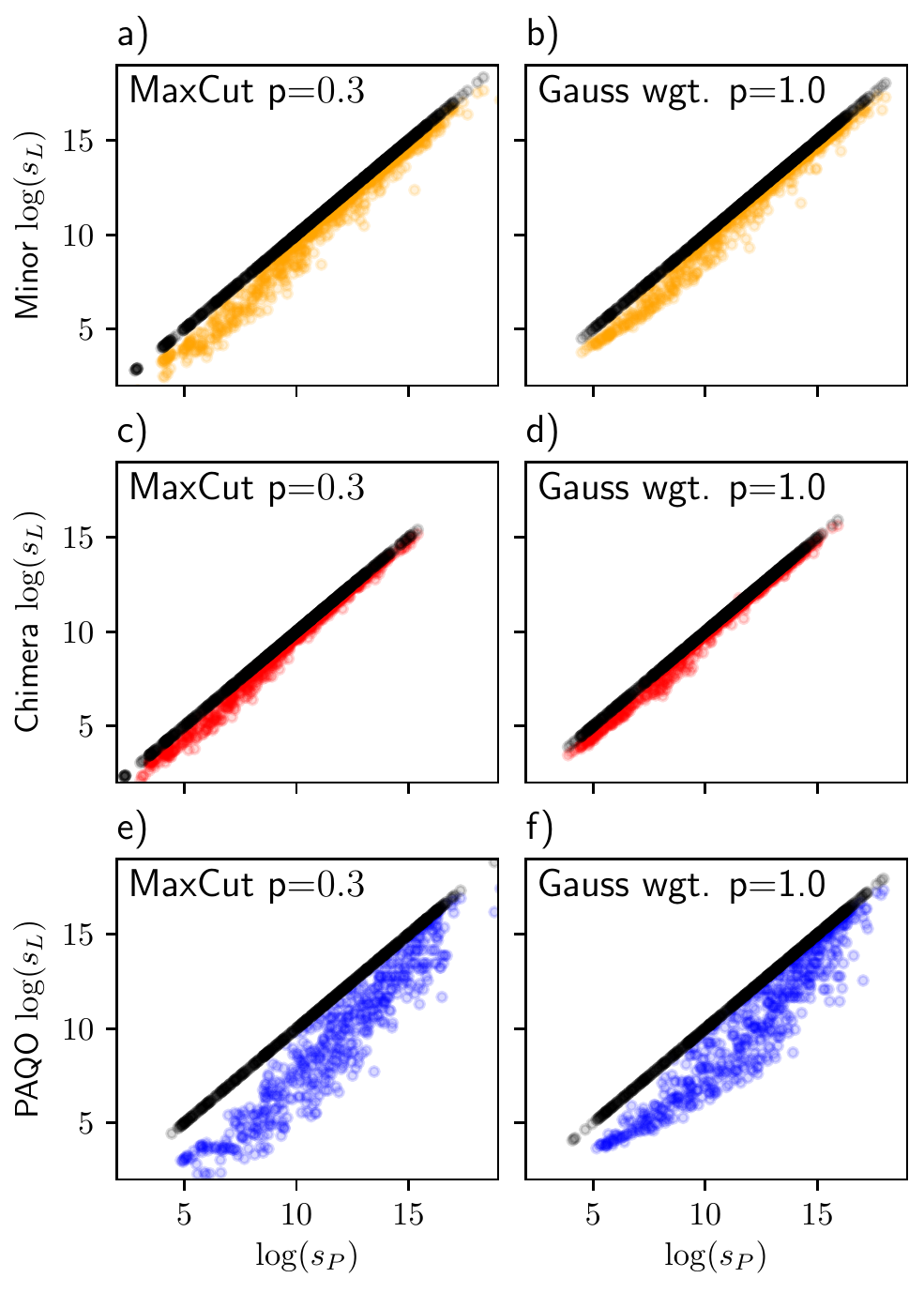}
\caption{\label{f_tts_vs_tts_decode}
$s_L$ plotted versus $s_P$ for the three embedding schemes and MaxCut ($p = 0.3$, left column) and Gaussian spin glasses (right column). A decoder can only make $s_L$ faster. If we run a trivial decoder that accepts no errors, then $s_L$ is equal to $s_P$ --- shown as the black circles. A trend to smaller benefits toward larger sizes (i.e., larger $s_P$) is visible, which leads to slightly worse scaling when fitting the data. Furthermore, it is visible that the decoder for the PAQO scheme is more efficient than majority voting for embedding on both the square and chimera lattices. While it reduces $s_L$ by orders of magnitude compared to $s_P$---which could be invaluable for any real-world applications---it does not improve the scaling.
}
\end{figure}

Fitting the data in Fig.~\ref{f_tts_vs_tts_separate_scatter} to a linear function of the form $y=m x+c$ yields the parameters shown in Tab.~\ref{t_fitting_parameter}. We show the fit for both $s_L$ and $s_P$. While it seems nonintuitive that $s_L$ scales worse than $s_P$, Fig.~\ref{f_tts_vs_tts_decode} shows that the decoding yields more benefits for small sizes and hence distorts the scaling. However, the actual graph of $s_L$ is bounded by $s_P$, because $s_P$ is trivially decodable. Therefore, the scaling of $s_L$ is also bounded eventually by $s_P$. Thus we assume the smaller scaling of the two. A stronger decoder might lead to a better scaling but most likely at a corresponding level of complexity regarding its run time.

\begin{table}
\caption{\label{t_fitting_parameter}The fitting parameters for a linear fitting function of the form $y=m x+c$ over the medians shown in Fig.~\ref{f_tts_vs_tts_separate_scatter}. Easier instances might scale better, while harder instance families might scale considerably worse. We include the fit parameters for $s_L$ and $s_P$, together with the correlation coefficient $r$ of the linear regression.}
\begin{tabular*}{\columnwidth}{@{\extracolsep{\fill}}llcccccr}
\hline\hline
Algorithm & Problem type & $m_L$ & $c_L$ & $r_L$ & $m_P$ & $c_P$ & $r_P$\\
\hline\hline
Square & MaxCut p=0.3 & 3.27 & -2.28 & 1.00  & 2.98 & -0.52 & 0.99 \\
Square & MaxCut p=0.5 & 3.45 & -2.95 & 0.99  & 3.06 & -0.66 & 0.99 \\
Square & Gaussian weights p=1.0 & 3.00 & -2.92 & 0.99  & 2.70 & -0.97 & 0.99 \\
\hline
Chimera & MaxCut p=0.3 & 2.73 & -1.59 & 1.00  & 2.55 & -0.57 & 0.99 \\
Chimera & MaxCut p=0.5 & 2.81 & -1.72 & 0.99  & 2.57 & -0.39 & 0.99 \\
Chimera & Gaussian weights p=1.0 & 2.54 & -1.95 & 0.99  & 2.35 & -0.84 & 0.99 \\
\hline
PAQO & MaxCut p=0.3 & 5.03 & -7.61 & 0.99  & 3.89 & -1.66 & 0.99 \\
PAQO & MaxCut p=0.5 & 4.31 & -6.05 & 0.99  & 3.80 & -1.23 & 0.99 \\
PAQO & Gaussian weights p=1.0 & 4.09 & -7.62 & 0.98  & 3.35 & -2.71 & 0.99 \\
\hline\hline
\end{tabular*}
\end{table}

\section{A note on Embedding penalties}
\label{sec:emb}

Both an all-to-all spin glass and the encoded spin glass suffer from an exponentially closing gap \cite{knysh:15}. Due to the chainlike nature of the physical spins representing a logical spin, all three embeddings are subject to the fact that the gap of a ferromagnetic chain in a weak field is exponentially suppressed \cite{isakov:16} with system size $N$. More precisely, Ref.~\cite{isakov:16} has shown the gap to be $\Delta \approx \Gamma^N/J^{N-1}$, where $\Gamma$ is the field strength and $J$ the ferromagnetic coupling of the chain. The gap is related to the tunneling rate to flip the chain, which determines the dynamics and hence the required annealing time $T$. The reason why the physical spin chain must be able to flip is motivated in Appendix~\ref{apx:avoid} with a three-spin example. Furthermore, the constraint-to-logical coupling ratio worsens with growing chains, because longer chains require stronger constraints to prevent breaks. However, an embedded system using $\mathcal{O}({N^2})$ spins is not as slow as a native $\mathcal{O}({N^2})$ problem, because the constraint structure is ferromagnetic, which is easier to solve.

\section{CONCLUSIONS}
\label{sec:conclusion}

We study the time overhead when solving quadratic binary optimization problems using quantum annealing after embedding the problems with three different embedding schemes. In all cases, there is a sizable time overhead in the time to solution with exponents between $m=2.35$ and $m=5.03$, and an exponential slowdown with respect to the system size $N$. A constant-speed advantage of analog quantum annealers is quickly decreased and turned into a slowdown by this run-time penalty for embedding. Therefore, in the absence of long-range physical couplers that would allow for the study of problems without the need of embedding schemes in superconducting quantum annealing machines, simulated quantum annealing scales better in time to solution than analog quantum annealers with quasiplanar graph topologies.
Similarly, programmable gate-model quantum computers could also implement arbitrary problems with arbitrary connectivity with only a polynomial and not an exponential overhead in the problem size $N$.

In light of these results, as far as analog quantum annealers are concerned, it is also of great interest to develop novel approaches to potentially overcome these limitations. These could include variable connectivity \cite{hauke2015probing}, nonlinear driving \cite{susa2020performance}, variational quantum annealing \cite{susa2021variational}, as well as counterdiabatic driving \cite{hartmann2019rapid}. The present results could serve as a benchmark for future developments of hardware quantum annealers.

\begin{acknowledgments}

M.S.K. thanks many members of the former Troyer Group at ETH Zurich for the fruitful discussions and inputs. The large-scale simulated quantum annealing calculations were done using the Daint Cluster at the Swiss National Supercomputing Center (CSCS). H.G.K. would like to thank Coco Gatograbber for multiple exchanges. W.L. was supported by the Austrian Science Fund (FWF) through a START grant under Project No. Y1067-N27, the Special Research Program(SFB) Information Systems beyond Classical Capabilities (BeyondC) Project No. F7108-N38, the Hauser-Raspe Foundation, and the European Union's Horizon 2020 research and innovation program, under Grant Agreement No. 817482. This material is based upon work supported by the Defense Advanced Research Projects Agency (DARPA) under Contract No. HR001120C0068. Any opinions, findings and conclusions or recommendations expressed in this material are those of the author(s) and do not necessarily reflect the views of DARPA.

\end{acknowledgments}

\appendix

\begin{figure}
\centering
\includegraphics[width=1\columnwidth]{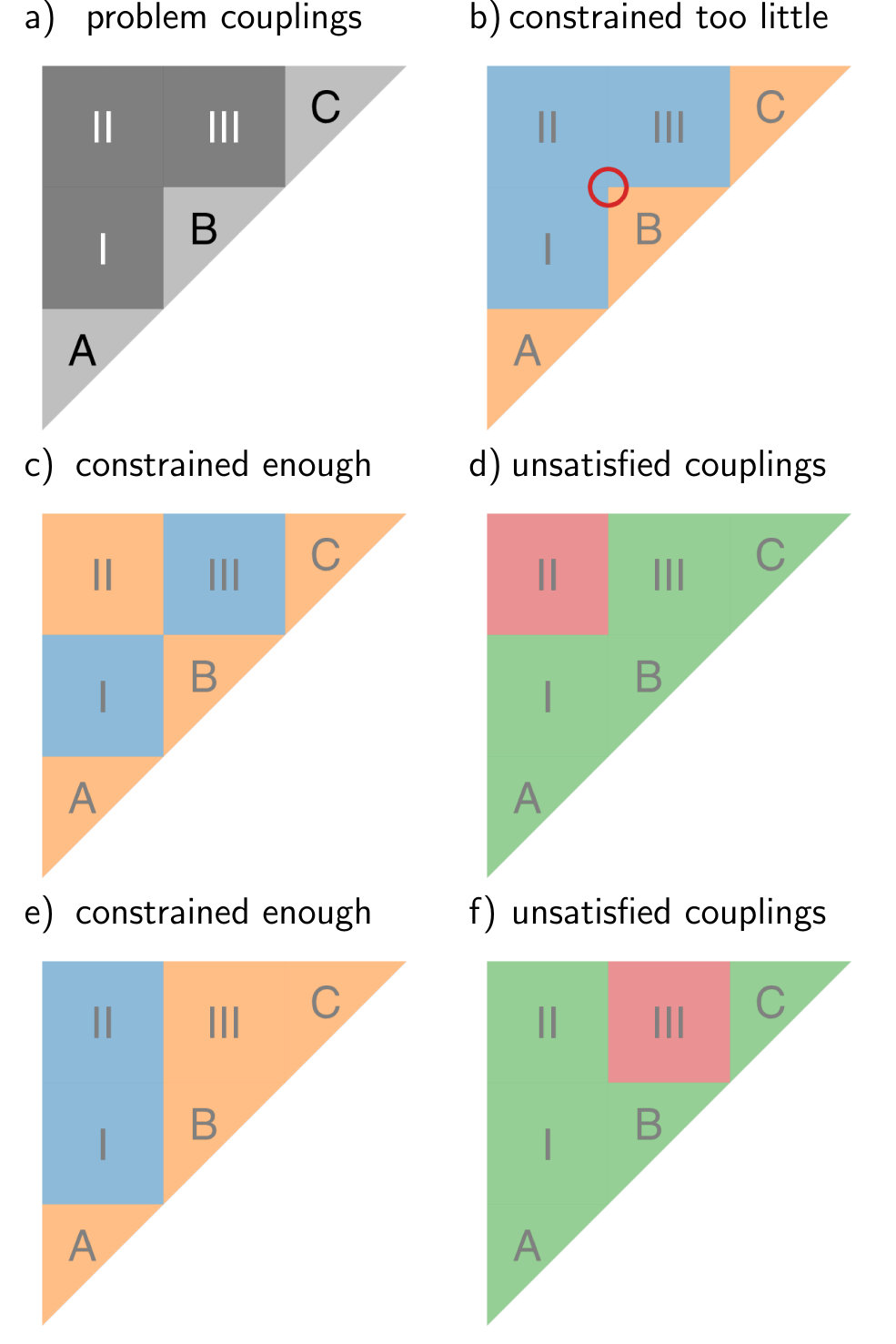}
\caption{\label{f_paqo_theo}
A constraint-scaling example in the PAQO scheme. The logical system is not shown. (a) The logical couplers are set up in three equal groups for a logical system of large size N; hence we do not display the individual physical spins as in Figure \ref{f_mappings}(c). All spins in A want to be parallel (light gray, $J=1$) to each other, as do the ones in B and C. Region I wants all A spins to be antiparallel (dark gray, $J=-1$) to all B spins. Region II is the same for all A and C spins and region III is the same for all B and C spins. (b) The physical ground state (blue down spin, orange up spin) if the $4$-local term constraint marked with the (red) circle is too weak. This broken ground state satisfies all problem couplings but does not correspond to a logical state. (c) If the constraints are strong enough, at least one of the regions is not satisfied [here, region II of area $(1/9)N^2$]. (d) Unsatisfied couplings from the solution in (c), where the red color marks physical spins that add energy. (e) Another possible physical ground state if constrained enough, with (f) showing the unsatisfied couplings thereof in region III, again of area $(1/9)N^2$.}
\end{figure}

\section{PAQO CONSTRAINTS}
\label{apx:paqo_constr}

\begin{figure}
\centering
\includegraphics[width=1\columnwidth]{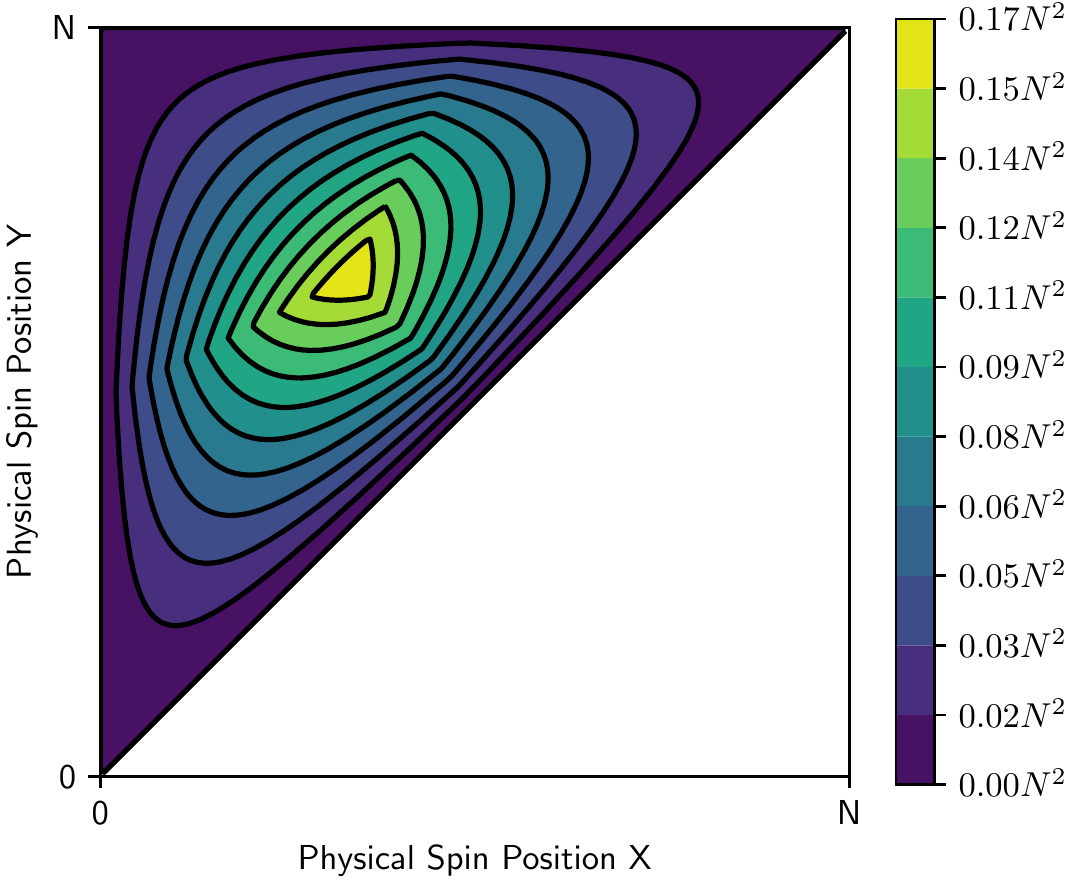}
\caption{\label{f_paqo_min_constr}
The minimal $4$-local-term constraint strength required. Using the example from Fig.~\ref{f_paqo_theo} and changing the number of spins in Groups A, B, and C (and therefore the size of the blocks I, II, and III) allows one to move the constraint position that breaks to any point and observe at which strength it breaks. This is mapped out in this contour plot in terms of $N^2$. It shows that the $4$-local-term constraint must be stronger in the middle of the PAQO embedding and can be weaker at toward the borders. This is, however, just one type of problem family and not a rigorous lower bound.}
\end{figure}

The constraints for the PAQO embedding need to be selected more carefully because a safe lower bound for the constraint strength is not immediately evident. Consider the following logical problem, which consists of $N$ all-to-all-connected spins divided into three groups, A, B, and C. Within each group, they are coupled with $J=1$ and want to be parallel. Each group is connected with $J=-1$ to any spin of the other groups, i.e. groups A, B, and C all want to be antiparallel to each other, making them frustrated amongst each other. Because it is not possible for A, B, and C to be all antiparallel, at least one pairing needs to be parallel. Figure \ref{f_paqo_theo} shows the corresponding PAQO embedding where allowing one broken constraint lowers the energy by $E=(1/9)N^2$. Note that this argument can be repeated in a recursive way in the regions A, B, and C to reach a bound of $E=(1/6)N^2$. This lower bound dictates the constraint to grow at least as fast to prevent being broken in the physical ground state. Because this is just one example, it is not a safe theoretical lower bound for any arbitrary problem. As such, calculating the necessary minimal constraints is more difficult than for square- and chimera-lattice embedding. We expand the analysis of Fig.~\ref{f_paqo_theo} in Fig.~\ref{f_paqo_min_constr}, where we change the problem in Fig.~\ref{f_paqo_theo}(a) such that we can observe the break at every point and plot the strength required to prevent it. Again, this is one problem family with one antiferromagnetic and one ferromagnetic coupling and not a guaranteed lower bound. But if any one constraint is below this bound, we can construct an instance for which the constraints are too weak to find the proper physical ground state. The decoder might still recover from this error, but it is straightforward to construct problems where this is not the case.

\section{A Note on Avoided Level Crossings}
\label{apx:avoid}

\begin{figure}
\centering
\includegraphics[width=1\columnwidth]{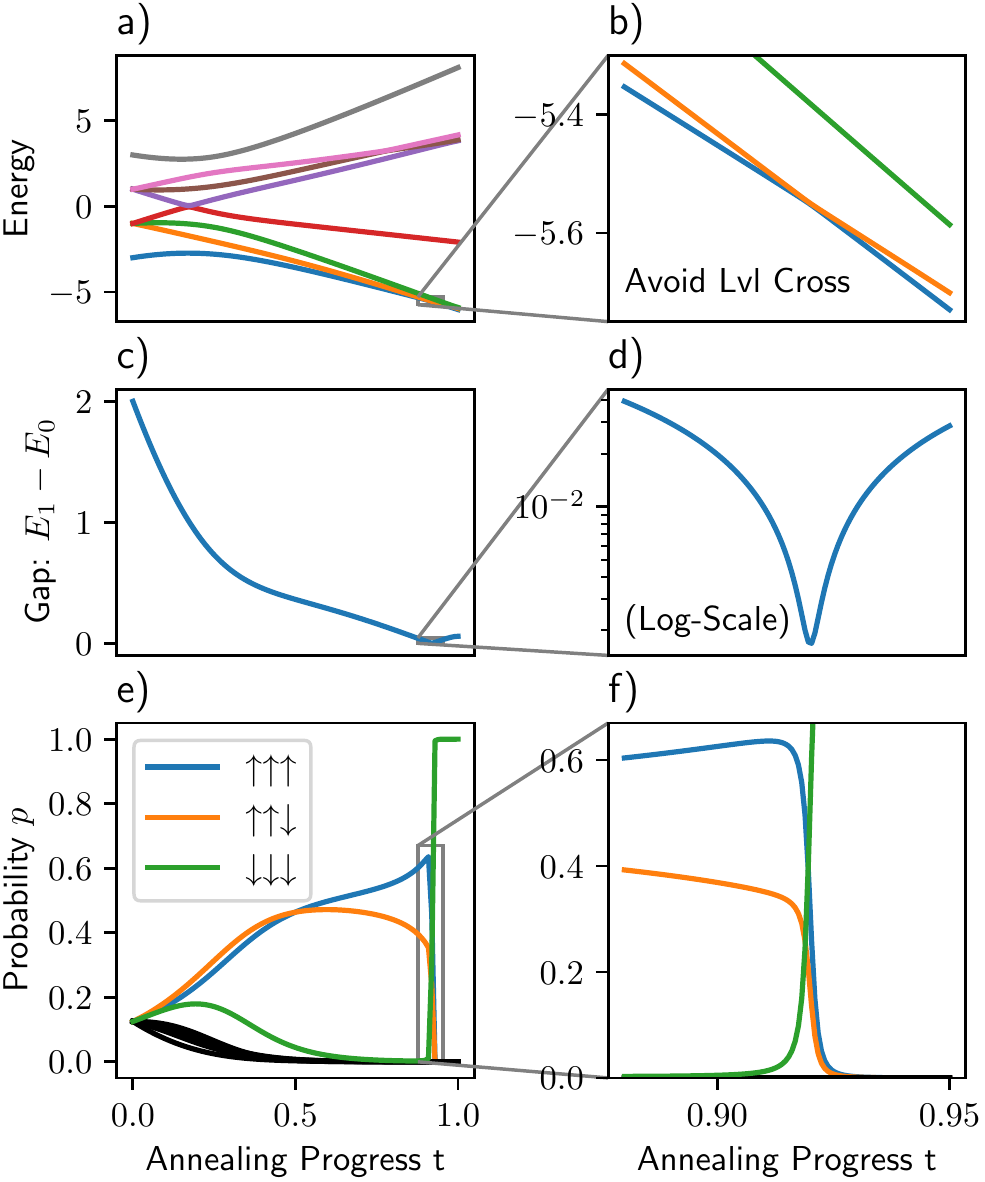}
\caption{\label{f_avoided_lvl_cross}
Spectrum, energy gap, and instantaneous ground state for a 3-spin ferromagnetic chain with individual local fields \cite{comment:fn4}. Panel (a) shows the spectrum of transverse field annealing, starting with the pure transverse field at $t=0$ and transforming to the pure problem Hamiltonian at $t=1$ in a linear manner. (b) Avoided level crossing between the two lowest states. Panel (c) show the first energy gap with a zoom in log scale shown in panel (d). The gap reaches $\Delta \approx 10^{-3}$ at its smallest position. (e) The instantaneous ground state is shown by the probabilities in the computational basis, of which we only label the three dominant parts and display the remaining five as black lines. Panel (f) shows the sudden rise and fall of the ground-state components representing a flip in the first two spins, and changing the third spin from a superposition to up.
}
\end{figure}

In hard-optimization problems, avoided level crossings \cite{knysh:15} arise in the annealing spectrum because the driver Hamiltonian might initially favor certain configurations that become unfavorable once the driver is weak enough. To remain in the instantaneous ground state and follow the adiabatic anneal path, the spins have to flip. In contrast problems where the initial horizontal spins (due to the transverse field driver) only move toward up or down and do not need to flip during the anneal are much easier to solve . We include a three-spin example \cite{comment:fn4} with the spectrum shown in Figs.~\ref{f_avoided_lvl_cross}(a) and \ref{f_avoided_lvl_cross}(b) that displays such an avoided crossing at $t\approx 0.92$, where $T$ is the annealing time and $t$ is the progress in the annealing schedule. The gap in Figs.~\ref{f_avoided_lvl_cross}(c) and \ref{f_avoided_lvl_cross}(d) shrinks linearly to near zero and then grows again. The instantaneous ground state displayed in computational-basis probabilities in Figs.~\ref{f_avoided_lvl_cross}(e) and \ref{f_avoided_lvl_cross}(f) clearly shows that the first two spins flip from up ($p_{\text{up}}\approx 1$) to down ($p_{\text{down}}\approx 1$), while the third spin changes from a superposition to down. Figures \ref{f_mappings}(a) -- \ref{f_mappings}(d) show the spins that need to be flipped if one flips the logical spin marked C. For square- and chimera-lattice embedding, it corresponds to a ferromagnetic chain that grows linearly with the system size $N$. In the PAQO scheme, we also require a number of flips that grows linearly with $N$. The physical spins are connected by $4$-local terms and hence form a chain structure as well, as can be seen in Fig.~\ref{f_mappings}(c). Any embedding needs to accommodate the flips of logical spins during the anneal (ideally) in an efficient manner.

\bibliography{refs,comments}

\begin{thebibliography}{58}
\bibitem[{com({\natexlab{a}})}]{comment:fn1}
\bibinfo{note}{This constraint modeling can be improved. Constraints closer to
  the chain end can be smaller, since they need only be stronger than the
  weaker part of the chain.}
\bibitem[{com({\natexlab{b}})}]{comment:fn3}
\bibinfo{note}{In absence of a $4$-local coupling therm, these constraints can
  be broken down into $2$-local coupling terms by introducing further auxiliary
  spins.}
\bibitem[{com({\natexlab{c}})}]{comment:fn4}
\bibinfo{note}{The coupling for the 3-spin systems are (unspecified energy
  unit): $J_{1,2}=5,J_{2,3}=1$ and the local fields are:
  $h_1=-0.1,h_2=1.04,h_3=-0.97$.}
\end{thebibliography}

\end{document}